\documentclass[aps,pra,twocolumn,groupedaddress]{revtex4-1}

\usepackage{graphicx}
\usepackage{amsmath}

\begin{document}

\title{Suitability of linear quadrupole ion traps for large Coulomb crystals}
\author{D. A. Tabor}
\author{V. Rajagopal}
\author{Y-W. Lin}
\author{B. Odom}
\affiliation{Department of Physics and Astronomy, Northwestern University
\\ 2145 Sheridan Rd., Evanston, IL 60208}
\date{\today}

\begin{abstract}
Growing and studying large Coulomb crystals, composed of tens to hundreds of thousands of ions, in linear quadrupole ion traps presents new challenges for trap implementation.  We consider several trap designs, first comparing the total driven micromotion amplitude as a function of location within the trapping volume; total micromotion is an important point of comparison since it can limit crystal size by transfer of radiofrequency drive energy into thermal energy.  We also compare the axial component of micromotion, which leads to first-order Doppler shifts along the preferred spectroscopy axis in precision measurements on large Coulomb crystals.  Finally, we compare trapping potential anharmonicity, which can induce nonlinear resonance heating by shifting normal mode frequencies onto resonance as a crystal grows. We apply a non-deforming crystal approximation for simple calculation of these anharmonicity-induced shifts, allowing a straightforward estimation of when crystal growth can lead to excitation of different nonlinear heating resonances. In the anharmonicity point of comparison, we find significant differences between the trap designs, with an original rotated-endcap trap performing better than the conventional in-line endcap trap.
\end{abstract}
\maketitle

\section{Introduction}
\label{sec-introduction}

Ion trapping by radiofrequency (rf) electric fields was first demonstrated in 1954 by Paul in a trap with hyperbolic electrodes \cite{paul90}.  Attempts to maximize the field-free trapping volume led to the development of the linear Paul trap by Prestage in 1989 \cite{prestage89}.  Although less harmonic than hyperbolic traps, the field-free central axis of linear traps often make theirs the preferred geometry for experiments involving multiple laser-cooled ions, including precision spectroscopy \cite{rosenband05, koelemeij07}, quantum information processing \cite{haffner05, hume07}, and cavity quantum electrodynamics applications \cite{herskind09}.

When the thermal energy of a trapped ion cloud is cooled sufficiently below the repulsive Coulomb interaction energy, the ions settle into a Coulomb crystal with geometry defined by the trapping potential \cite{willitsch08, setsuo82}.  Crystallization of certain atomic ion species with closed-cycle electronic transitions can be accomplished directly by laser cooling \cite{drewsen98, removille09}; crystallization of other species (e.g., molecular ions) can be accomplished sympathetically by laser cooling of a co-trapped atomic species \cite{blythe05} or \emph{in situ} formation from a pre-cooled reactant \cite{molhave00}. Crystallization can be prevented by various heating mechanisms.  Two such heating mechanisms, \textit{micromotion heating} and \textit{nonlinear resonance heating}, become more problematic as a crystal grows, due to sampling of larger rf fields and of nonharmonic regions of the trapping potential, respectively. Thus, these heating mechanisms can limit the minimum temperature obtained for a given crystal size and also the maximum obtainable crystal size. Our work is motivated in part by recent difficulties growing large Coulomb crystals in a non-standard trap geometry, leading to speculation that axial rf fields near the endcaps (fringing fields) might limit crystal size \cite{kielpinski10}.

In this paper, we compare the suitability of four linear Paul trap designs for growth and study of large Coulomb crystals.  First, we compare the total micromotion (the unavoidable driven motion at the rf frequency) which results in non-resonant transfer of rf energy into thermal energy of the trapped sample (micromotion heating).  Second, we compare the component of micromotion along the trap axis (the axis of smallest micromotion), which induces first-order Doppler shifts in precision spectroscopy on large Coulomb crystals. Finally, we compare trap anharmonicity, which can lead to excitation of nonlinear heating resonances as a crystal grows and its normal mode frequencies are shifted onto a resonance. To characterize these anharmonicity-induced frequency shifts, we use a simple non-deforming crystal approximation, which allows a straightforward estimation of when crystal growth can lead to excitation of different nonlinear heating resonances.

\section{Trap Designs}
\label{sec-trapdesigns}

\begin{figure}
\includegraphics[width=85mm]{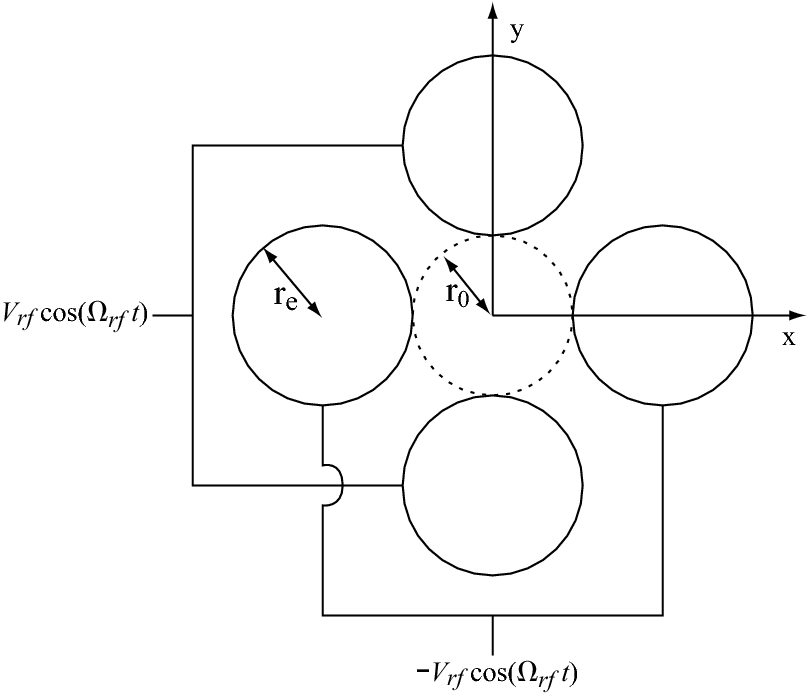}
\caption{\label{fig-trap} End view of a linear quadrupole trap.  A sinusoidally varying voltage is applied to the rod electrodes, with one pair held $180^\circ$ out of phase with the other.}
\end{figure}

A linear quadrupole ion trap (Fig.~\ref{fig-trap}) is formed from four parallel cylindrical electrodes of radius $r_e$, held at a separation $r_0$, with an rf voltage signal applied to the electrodes.  The resulting electric field confines the ion motion radially; axial confinement (in the $z$-direction) is provided by endcap electrodes (not shown in Fig.~\ref{fig-trap}), to which a static voltage is applied.

Near the trap center, the electric potential is given by
\begin{eqnarray} 
\label{eq-potlatcenter}
   \phi (x,y,z,t) &=& \frac{V_{rf}cos(\Omega_{rf}t)}{r_0^2}(x^2-y^2) + \nonumber \\
   && \frac{\kappa V_{ec}}{z_0^2}(\frac{2z^2-x^2-y^2}{2})
\end{eqnarray}
where $V_{rf}$ and $\Omega_{rf}$ are the voltage and frequency of the applied rf drive, $V_{ec}$ is the static voltage applied to the endcap electrodes, $z_0$ is half the distance between the endcap electrodes, and $\kappa$ is a geometric factor.  While this description is accurate near the trap center and along the $z$-axis, significant deviations can occur elsewhere in the trap interior.

The radial motion of an ion in the potential of Eq.~\eqref{eq-potlatcenter} is described by the Mathieu differential equations, and the stability of this motion is expressed using Mathieu parameters which depend only on $V_{rf}$, $V_{ec}$, $\Omega_{rf}$, $\kappa$, and the charge-to-mass ratio $q/m$.  Solutions to the Mathieu equations contain regions of parameter space where ion motion is stable \cite{mclachlan47}, and trap parameters are chosen to operate in one such stable region.

The motion of a trapped ion can be described approximately as a superposition of micromotion, in which the ion oscillates at the rf frequency, with secular motion, in which the ion oscillates in a time-independent pseudopotential $\tilde{\phi}_{rf}(x,y,z)$ at slower secular frequencies $\omega_x$, $\omega_y$, and $\omega_z$.   An analytic relationship exists between the time-dependent rf potential and the time-independent pseudopotential \cite{dehmelt67}:
\begin{equation}
\tilde{\phi}_{rf}(x,y,z) =\frac{q}{4m\Omega_{rf}^2} |\nabla{}\phi_{rf}(x,y,z,t=0)|^2
\end{equation}
The endcap electrodes generate a potential $\phi_{ec}$, and the total potential governing secular motion is
\begin{equation}
\phi_{trap}=\tilde{\phi}_{rf}+\phi_{ec}.
\end{equation}

\begin{figure}
\includegraphics[width=85mm]{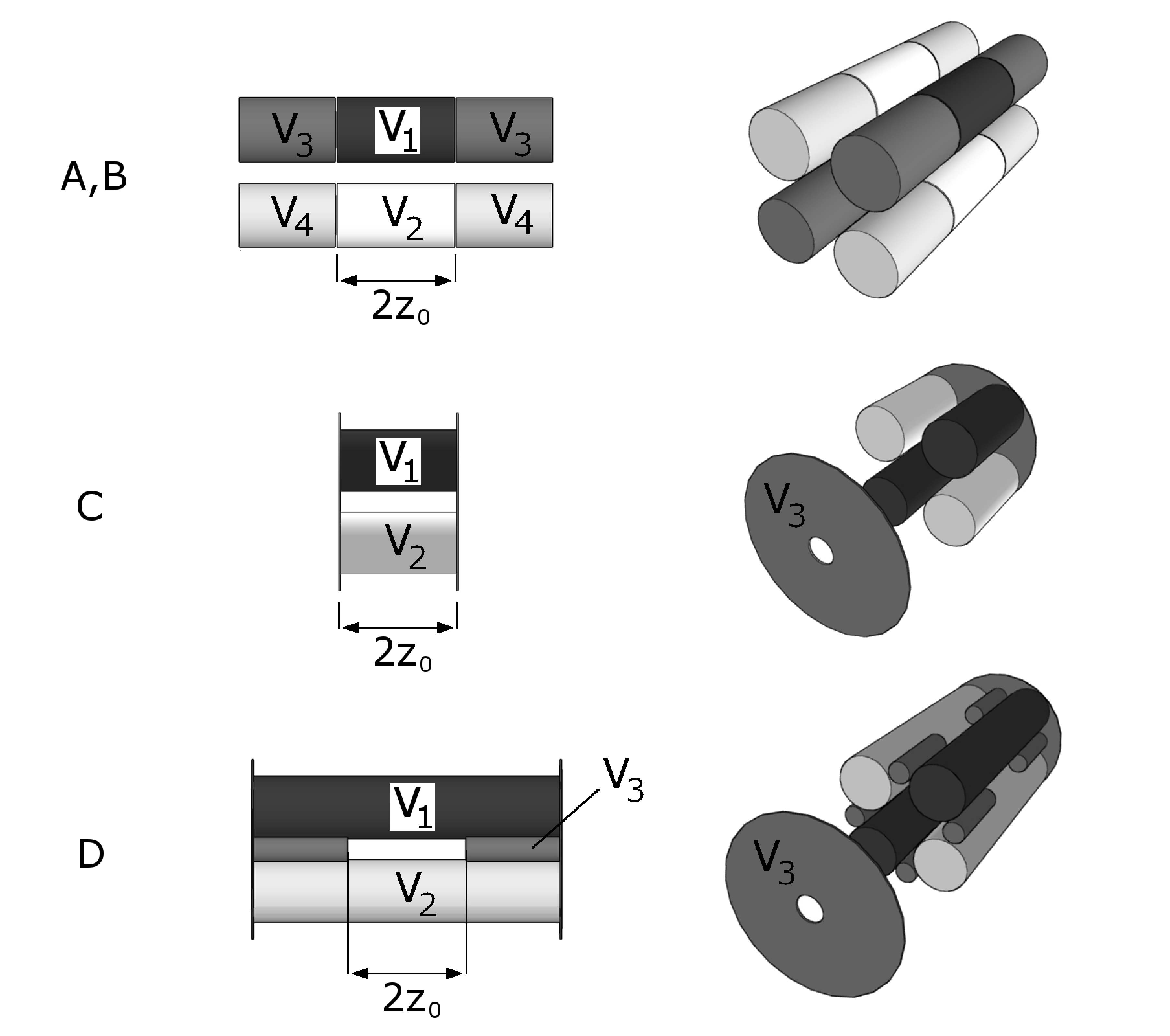}
\caption{\label{fig-trapgeoms} Trap designs analyzed in this
paper.  Designs $A$ and $B$ share the same geometry, but differ in applied voltages.  In the offset view at right, designs $C$ and $D$ are shown with the nearest plate endcap pulled away from the rest of the trap to allow the trap interior to be seen. }
\end{figure}

We compare four different linear trap implementations (Fig.~\ref{fig-trapgeoms}, with corresponding voltage given in Table~\ref{tbl-voltages}): a conventional in-line endcap design with two different voltage configurations (designs $A$ and $B$), a plate endcap design (design $C$, similar to a design analyzed in Ref. \cite{pedregosa10}), and an original rotated endcap design (design $D$).  In all trap designs, $z_0$ = 8.5 mm,  $r_0$ = 4 mm, and $r_e$ = 4.5 mm, with $r_0$ and $r_e$ chosen to agree with the optimal ratio $r_e$=1.14511$r_0$ for minimizing the leading-order contribution to anharmonicity in the rf potential \cite{reuben96}. Design $C$ uses thin circular plates of radius $2r_e+r_0$ as endcap electrodes, with a 2 mm radius hole left for axial optical access.  Design $D$ has four small cylindrical endcap electrodes of radius 1.75 mm, rotated 45$^{\circ}$ around the $z$-axis; the central axes of these endcap electrodes are on the same radius as the central axes of the rf electrodes. Our group has built a functional trap whose key geometric features are shared with design $D$, currently being used to trap $^{138}$Ba$^{+}$ ions.  All numerical simulations in this paper use values corresponding to $^{138}$Ba$^{+}$  ($q=1.602\times10^{-19}$ C, $m=2.292\times10^{-25}$ kg) and our operating rf frequency ($\Omega_{rf}=2\pi\times3.00$ MHz).

\begin{table}
\caption{\label{tbl-voltages} Operating voltages applied in different trap designs. Here $V(t)=+V_{rf}cos(\Omega_{rf}t)$.}
\begin{ruledtabular}
\begin{tabular}{ c  c  c  c  c }
 & $V_1$ & $V_2$  & $V_3$ & $V_4$\\
 \hline
 $A$ & $V(t)$ & $-V(t)$ & $V_{ec}$ & $V_{ec}$ \\
 $B$ & $V(t)$ & $-V(t)$ & $V_{ec}+V(t)$ & $V_{ec}-V(t)$ \\
 $C$ & $V(t)$ & $-V(t)$ & $V_{ec}$ & $N/A$ \\
 $D$ & $V(t)$ & $-V(t)$ & $V_{ec}$ & $N/A$ \\
\end{tabular}
\end{ruledtabular}
\end{table}

\section{Computation of Trapping Potentials}
\label{sec-compmethod}

Our metrics of trap design comparison, presented in Sec.~\ref{sec-mmotion} and Sec.~\ref{sec-comfrequency}, are calculated from the trap potentials and their gradients.  The potentials $\phi_{ec}$ and $\phi_{rf}$ are found using the finite element method to solve the Laplace equation, with the trap geometries discretized by the meshing software Gmsh \cite{geuzaine09}.  

To allow evaluation of the potentials at locations not on the mesh, the numerical solutions are fitted to an expansion in associated Legendre polynomials
\begin{eqnarray}
\label{alpseries}
\phi_{fit}(r, \theta, \phi) = \sum_{l=0}^{l_{max}} \ \sum_{m=-l_{max}}^{l_{max}} C_{lm}\ r^l\ P_{l}^m(cos~\theta) \nonumber \\
   \times
   \begin{cases}
      cos(m\phi) & \text{for $m \ge 0$} \\
      sin(m\phi) & \text{for $m < 0$}.
   \end{cases}
\end{eqnarray}
We exclude terms with odd values of $l$ and $m$ from our fitting due to symmetry.

An analytic form of the gradient of Eq.~\eqref{alpseries} is straightforward to find using
\begin{eqnarray}
\label{alpgrad}
\lefteqn{ \frac{\partial P_{l}^m(cos~\theta)}{\partial \theta} = }  \nonumber \\
&& \frac{l \: cos\, \theta \: P_{l}^m(cos~\theta) - (l+m) P_{-l}^m(cos~\theta))}{\sqrt{1-cos^2\theta}},
\end{eqnarray}
which can be derived from a more general identity given by \cite{abramowitz64}.  Fitting the Laplace solutions to the above expansion thus permits efficient evaluation of $\phi_{fit}$ or its gradient $\nabla \phi_{fit}$ at arbitrary location from the coefficients $\{C_{lm}\}$.  In this way, we determine $\tilde{\phi}_{rf}$ and $\phi_{ec}$ up to the scaling set by choosing $V_{rf}$ and $V_{ec}$.  Our choices of these voltages, shown in Table~\ref{tbl-kappas}, are made to equalize the single particle secular frequencies $\omega_z$ and $\omega_r$ across all designs at experimentally reasonable values $\omega_x=\omega_y=2\pi\times418$ kHz and $\omega_z=2\pi\times18.9$ kHz.

Single particle secular frequencies are calculated by evaluation of $\tilde{\phi}_{rf}$ and $\phi_{ec}$ along a radial or axial trace through the origin from $r$=$-\frac{r_0}{2}$ to $r$=$+\frac{r_0}{2}$ or $z$=$-\frac{z_0}{2}$ to $z$=$\frac{z_0}{2}$ respectively.  The resulting values fit a polynomial whose quadratic coefficient rapidly converges with finer discretization of the domain or increase of the polynomial fitting order.  From this coefficient $C_2$, we also find the values of $\kappa=\sqrt{2qC_2/m}$ in Table~\ref{tbl-kappas}.

\begin{table}
\caption{$V_{rf}$ and $V_{ec}$ used in comparisons and calculated value of $\kappa$ for each trap.}
\begin{ruledtabular}
\begin{tabular}{ c c c c }
 Trap Design & $V_{rf}$ (volts) & $V_{ec}$ (volts) & $\kappa$ \\
 \hline
 \ A & 800 & 5.0 & 0.15 \\
 \ B & 800 & 5.0 & 0.15 \\
 \ C & 800 & 2.7 & 0.27 \\
 \ D & 800 & 51 & 0.014 \\
\end{tabular}
\end{ruledtabular}
\label{tbl-kappas}
\end{table}

The effect of inaccuracy in the Laplace solutions is estimated by observing convergence of calculated values such as $\kappa$ with increasing meshing density, which is found to cause variation in all reported results of less than a few percent unless otherwise noted as we approach the highest mesh densities eventually used -- of order $10^6$ nodes for all designs.  We discuss the implications of the resulting uncertainty in Sec.~\ref{sec-discussion}.  Tuning of other parameters, such as the fitting order in Eq.~\eqref{alpseries}, is found to cause negligible variance by comparison.

\section{Micromotion}
\label{sec-mmotion}

Although radial and axial micromotion vanish, respectively, along the trap axis and at $z=0$, a large crystal necessarily extends beyond these regions. Micromotion heating results from the transfer of micromotion energy into secular energy, which leads to an elevated secular temperature. Although the scaling is non-trivial, simulations show that the micromotion heating rate strongly increases with micromotion amplitude and with secular temperature \cite{schiffer00, ryjkov05}. As a crystal grows, additional ions are held at locations with increasingly strong rf fields, causing them to experience increasingly large micromotion. The resulting heating limits the minimum obtainable temperature for a crystal of a given size, and eventually prevents further crystal growth.

The electric field created by the rf electrodes may be written as
\begin{equation}
\overrightarrow{\mathbf{E}}(x,y,z,t) = \overrightarrow{\mathbf{E}}_\mathbf{0}(x,y,z) cos(\Omega_{rf} t).
\end{equation}
For an ion held in place in a stationary crystal, in the limit of small micromotion amplitude $\overrightarrow{\mathbf{E}}_\mathbf{0}$ may be taken as constant as the ion moves, and the micromotion of the ion is harmonic with amplitude
\begin{equation}
A(x,y,z)=\frac{q}{m\Omega_{rf}^2}|\nabla \phi_{rf}(x,y,z,t=0)|
\end{equation}

\begin{figure}
\includegraphics[width=85mm]{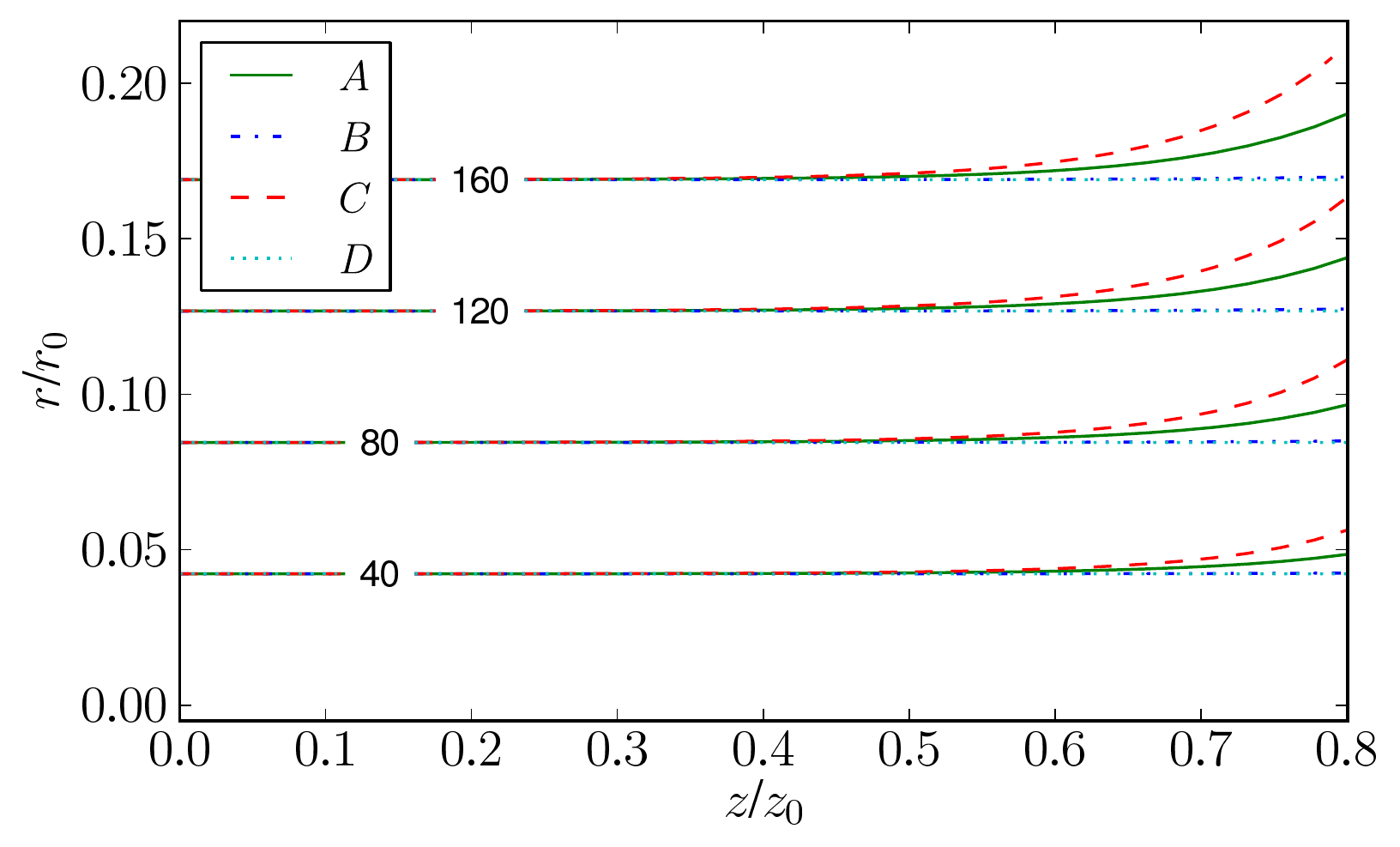}
\caption{\label{fig-totmmotion} (Color online.) Contour plots of total micromotion amplitude, labeled in microns. Designs $B$ and $D$ are indistinguishable at this scale.}
\end{figure}

With $V_{rf}$ as set in Table~\ref{tbl-kappas}, we find $\phi_{rf}(x,y,z,t=0)$ and calculate the resulting total micromotion amplitude $A(x,y,z)$.  The resulting micromotion contour is plotted in Fig~\ref{fig-totmmotion}.  All designs are comparable in the amplitude of total micromotion for any realistic trapping volume.  Interpretation of this result is discussed in Sec.~\ref{sec-discussion}.

We also compare the size of the $\hat{z}$ component of micromotion in each trap, which can limit the precision of spectroscopy.  Due to the smallness of these values (several orders of magnitude smaller than the total amplitude) numerical convergence comparable to Fig.~\ref{fig-totmmotion} is not observed.  Rather than plot contours, we compute the root mean square (RMS) average over a large trapping volume for each design in Table~\ref{tbl-axialmmotion}, producing a single quantity whose convergence is verified.  See Sec.~\ref{sec-discussion} for discussion.

\begin{table}
\caption{RMS average axial micromotion amplitudes over the cylindrical volume $r<500\mu$m, $|z|<z_0=8.5$mm.  Listed values for $A$ and $C$ change by no more than several percent when Laplace solutions are calculated over an order of magnitude in grid density.  Values for $B$ and $D$ change by up to tens of percent.  For averaging, the volume is discretized by a Cartesian grid with $h=20\mu$m, which is verified to be sufficiently dense.}
\begin{ruledtabular}
\begin{tabular}{ c c }
 Trap Design &  RMS Axial Micromotion Amplitude ($\mu$m)\\
 \hline
 \ A & 1.5 \\
 \ B & 0.04 \\
 \ C & 2.9 \\
 \ D & 0.08 \\
\end{tabular}
\end{ruledtabular}
\label{tbl-axialmmotion}
\end{table}

\section{Nonlinear Resonance Heating}
\label{sec-nlr}

Nonlinear resonance heating can occur if a resonance condition is satisfied between secular frequencies and the rf drive frequency.  For linear Paul traps, the condition is
\begin{equation}\label{eq-nlrcondition}
n_x\omega_x + n_y\omega_y + n_z\omega_z = \Omega_{rf}
\end{equation}
where $n_x$, $n_y$, and $n_z$ are integers \cite{drakoudis06}.  The corresponding condition for a single particle in a hyperbolic trap has been derived \cite{wang93}, with the resonances weakening with larger $n$.  This behavior has been confirmed in both hyperbolic traps \cite{alheit96} and linear traps \cite{drakoudis06}.  For sufficiently cold trapped samples, it is understood that the frequency of the center of mass (COM) modes, or other normal modes, should be used in Eq.~\eqref{eq-nlrcondition} rather than the single-particle frequencies \cite{wineland98}.  For an infinitely long linear Paul trap, or for a small crystal in a finite trap, where the axial rf drive vanishes, a non-deforming crystal is only expected to be heated on resonances with $n_z=0$.  However, a crystal which is large enough to sample axial fringing fields near the endcaps is expected to be excited also by resonances with $n_z > 0$.

Anharmonicity in the trapping potential causes the mode frequencies to shift as a crystal grows and samples less harmonic regions of the trap potential.  Eventually, the anharmonicity-shifted mode frequencies of a growing crystal will meet the resonance condition of Eq.~\eqref{eq-nlrcondition}, and some level of heating will occur, potentially halting further crystal growth. Even for a crystal of definite size, it is non-trivial to predict the precise response when it is swept through a given nonlinear heating resonance.  Predicting the heating response in the non-equilibrium scenario of a growing crystal, as anharmonic frequency shifts cause it to cross a resonance, is yet more complicated. Modeling of the heating rates and whether crossing a given subharmonic resonance will actually melt a crystal or prevent further growth is beyond the scope of this analysis.

Since it is difficult to determine whether crossing a heating resonance will limit crystal growth or cause other deleterious effects, in the current analysis, we take a conservative approach of attempting to design a trap in which a growing crystal will avoid low-order resonances altogether. We consider the anharmonicity-induced shifts of the COM frequencies for a non-deforming crystal; the non-deforming approximation is exact in the zero-temperature limit. In light of the resonant condition of Eq.~\eqref{eq-nlrcondition}, considering the COM frequency shifts provides a simple estimate of when trap anharmonicity could limit crystal growth.  Other modes of higher order than the COM mode can also be excited by nonlinear resonance heating, but the COM shifts are easily calculable and set the scale for anharmonicity-induced shifts in higher modes.

A previous analysis by the Marseille group (Pedregosa et. al. \cite{pedregosa10}) specified a single quantity describing the radial trap anharmonicity averaged over a trapped sample.  The Marseille approach has the utility of providing a single figure of merit by which to compare different trap designs; however, it does not consider axial anharmonicity (potentially important for long crystals) or provide an estimate of what level of anharmonicity is acceptable in order to avoid heating resonances.  The analysis we present here is similarly simple to calculate, and at the cost of not providing a single figure of merit, seeks to include axial anharmonicity and to provide a mechanism for determining whether a trap is sufficiently harmonic to safely work with crystals of some size.

\section{COM Frequency}
\label{sec-comfrequency}

For a zero-temperature crystal in a perfectly harmonic trapping potential, the COM oscillation frequency is the same as the secular frequency of a single particle. However, if the temperature of the crystal is nonzero or if the trapping potential is anharmonic, the crystal deforms during oscillation,  and the COM frequency and single particle frequency are no longer equal.  (Space charge shifts, which are well-understood in the limit of each ion moving independently in a background potential describing the distributed charge of the other ions, occur at high temperatures but not for a non-deforming crystal.)  Molecular dynamics (MD) simulations can find the COM frequency under finite temperature conditions, but this approach is computationally intensive for a large crystal. To understand the effects of anharmonicity with less computational overhead, we introduce the approximation of a non-deforming crystal in an anharmonic trap.  This approach is computationally simple enough to be useful in designing traps for large Coulomb crystal experiments.

A single ion in a zero temperature crystal is held in place by the cancellation of two forces: that due to the trap potential $F_{trap}$ and that due to the Coulomb repulsion of all other ions in the trap $F_{ions}$.  In the approximation that the crystal does not deform during oscillation, $F_{ions}$ remains constant while $F_{ions}+F_{trap}$ is, to leading order, a restoring force resulting in simple harmonic motion.

$F_{trap}$ can be calculated from $\phi_{trap}$, evaluated on a grid $(x_i,y_j,z_k)$ where $x_i=ih, y_j=jh$, and $z_k=kh$; $i$, $j$, $k$ are integers; $h$=50$\mu$m is the step size used in our calculation.  To illustrate our approach, we discuss calculation of the restoring force for small displacements in the $\hat{z}$ direction, at fixed $x_i$ and $y_j$.  By geometrical symmetry $\phi_{trap}$ is an even function of $z$, so the numerically calculated values can be fitted to a polynomial series of even terms:
\begin{equation}
\phi_{trap} (z) \Big |_{x_i, y_j}=C_0+C_2z^2+...+C_{10}z^{10}
\end{equation}
The axial force on an ion in a potential of this form is
\begin{equation}\label{eq-ftrap}
F_{trap}(z)=-q( 2C_2z + ... + 10C_{10}z^9).
\end{equation}
We now consider an ion whose equilibrium location in a crystal is $z=z_k$.  An expansion of $F_{trap}$ about $z_k$ gives
\begin{equation}\label{eq-fexp}
F_{trap}(z)=F_{trap}(z_k)+(z-z_k)F'_{trap}(z_k)+...
\end{equation}
Since $z=z_k$ is an equilibrium position for this ion, the Coulomb force due to all other ions $F_{ions}$ must exactly cancel the first term of this expansion.  In the approximation of a non-deforming crystal, $F_{ions}$ is a constant, and the total force on the ion is
\begin{align}\label{eq-ftot}
F_{trap}+F_{ions}&=-q\left( 2C_2 + ... + 90C_{10}z_k^8 \right)\left( z-z_k \right)
\\ &\equiv -k_z\Big |_{x_i, y_j, z_k} (z-z_k),
\end{align}
where $k_z|_{x_i, y_j, z_k}$ is a local spring constant describing the strength of the axial restoring force on an ion about its equilibrium location $(x_i, y_j, z_k)$.  As seen from Eqs.~\eqref{eq-ftrap} and \eqref{eq-ftot}, when $z_k\not=0$, anharmonic terms of the trap potential contribute to the spring constant $k_{z_k}$.

The restoring force on a crystal is the sum of the restoring forces on the individual ions.  For a non-deforming crystal, the result is oscillatory motion at a frequency set by the averaged local spring constants, with the average taken over the ion locations.  In our calculation we average instead over the grid locations, for points interior to the crystal volume. In this way, we can estimate the axial COM frequency for a crystal, given $\phi_{trap}$ and a specified crystal geometry.  We compute the COM frequencies for radial motion by the same method, again using a fitting polynomial of the same order.

\begin{figure}
\includegraphics[width=85mm]{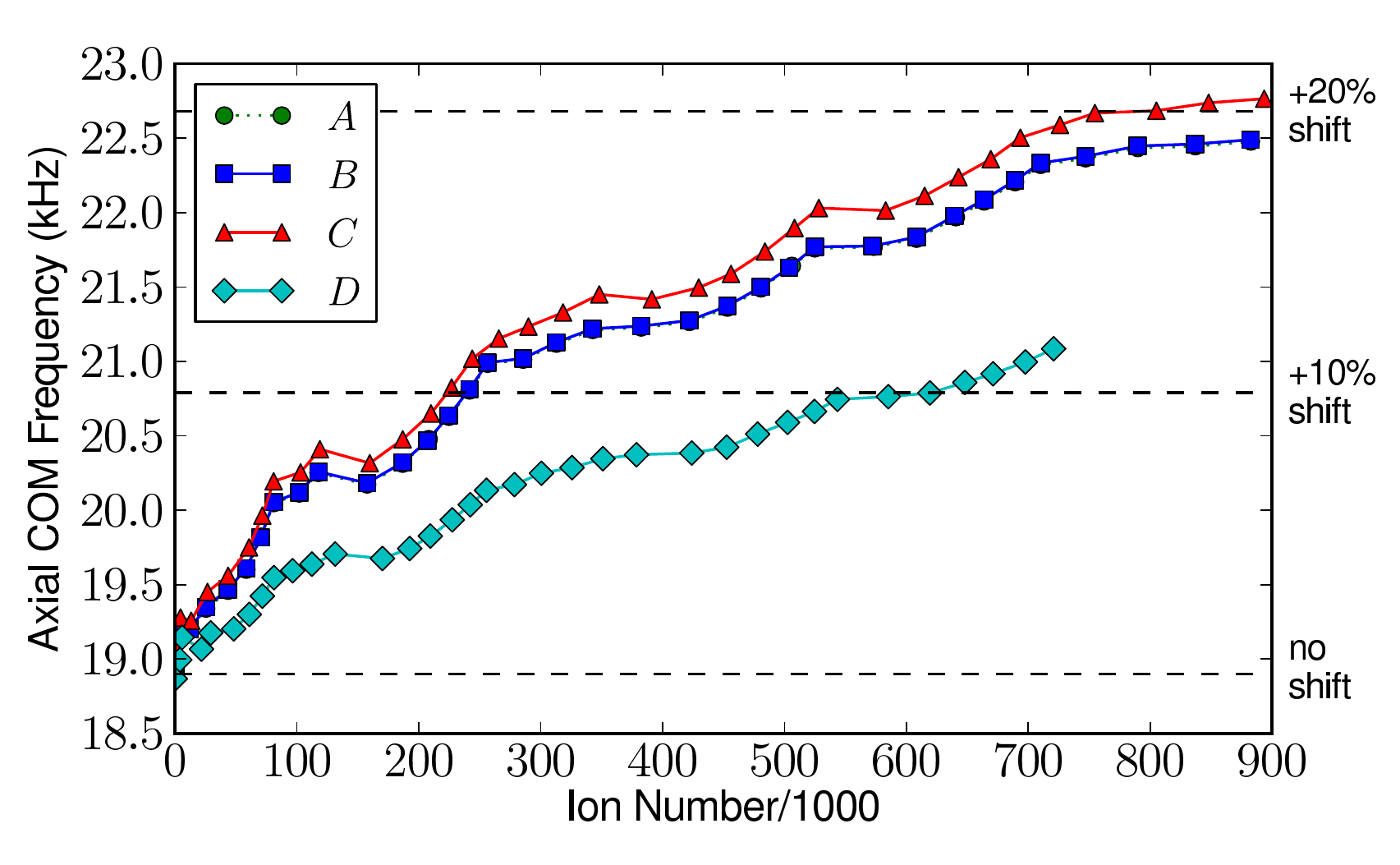}
\caption{\label{fig-wz} (Color online.) COM axial secular frequency as a function of ion number.  Curves for Designs $A$ and $B$ are overlapping.  Crystals are grown until their axial extent reaches ${\pm z_0/2}$; for equal length, design $D$ grows an appreciably smaller crystal.  Ripples result from finite evaluation grid size.  (See Sec.~\ref{sec-discussion} for discussion of both ion number in $D$ as well as finite grid effects.)}
\end{figure}

Fig.~\ref{fig-wz} shows the change in axial COM frequency in each trap design as crystal volume grows.  Crystal extent is determined by evaluation of $\phi_{trap}$ on a Cartesian grid with spacing $h$=50$\mu$m, with all points below a fixed cutoff included, and ion number is then determined by the ion density \cite{dubin99},  
\begin{equation}\label{eq-ion_density}
\rho = \frac{m\epsilon_0}{q^2}(\omega_z^2+2\omega_r^2),
\end{equation}
where $\omega_z$ and $\omega_r = \omega_x = \omega_y$ are the single particle secular frequencies.  We take this density as a constant; the anharmonicity-induced position dependence is negligible.  Implications of anharmonicity-induced frequency shifts for nonlinear resonance heating are discussed in Sec.\,\ref{sec-discussion}.

Radial COM shifts are also computed, with fittings made to radial rather than axial traces of $\phi_{trap}$.  Over the volume occupied by the largest crystals in Fig.~\ref{fig-wz}, no difference among the designs can be determined at our level of computational accuracy.  By varying all calculation parameters, we bound radial COM shifts at no more than tenths of a percent.

\section{Discussion}
\label{sec-discussion}

No difference in micromotion heating rate is expected among the different designs, since total micromotion amplitudes are comparable.  Indeed, in all cases axial micromotion amplitudes are much smaller than total amplitudes, indicating that rf-fringing should not limit crystal growth for these designs.  The actual limit on crystal size due to micromotion heating is difficult to estimate, but it can in principle be found from molecular dynamics simulations.

Axial micromotion, quantified by RMS averaging over a volume bounding any realistic crystal, is at least than an order of magnitude smaller  in designs $B$ and $D$ (in which the rf electrodes extend to $z > z_0$) than in $A$ and $C$ (in which the rf electrodes end at $z = z_0$).  Since axial micromotion results from fringing of the rf fields, these differences are expected, given the different rf boundary conditions at $z=\pm z_0$.  Given a conservative interpretation of the observed convergence of this average, $B$ and $D$ may be said to have comparable axial micromotion, while the situation in $A$ is much worse and in $C$ worse still.

The size of COM frequency shifts during crystal growth, found here to differ among the considered designs, is an indicator for the effect of nonlinear resonance heating on trap operation.  As the experimenter grows a crystal, adding ions continuously from an ion source, small COM frequency shifts (say a few percent) are not expected to present significant challenge.  The COM frequencies are unlikely to shift onto a nonlinear resonance during the intermediate stages of crystal growth, and the experimenter can load the largest crystal possible before checking if different voltages increase the maximum crystal size.  For larger size-dependent shifts, trapping voltages could, in principle, be dynamically tuned to avoid resonances as the crystal grows, but dynamic tuning would become increasingly challenging for larger shifts.  Thus the designs exhibiting large COM frequency shifts are expected to be more problematic for growing large crystals.  

As an example of these concerns, we note that for the frequencies considered in this study, excitation of the nonlinear resonance $|n_x|+|n_y|=7$, $n_z=3$ is expected when the axial COM frequency increases by 5-10\%, a shift which occurs in design $D$ only for a crystal containing tens of thousands more ions than in other designs.  (Resonances of this order and higher have been observed experimentally in a linear trap \cite{drakoudis06}.)  The expected ability of $D$ to load more ions than the other designs before exciting this resonance is a direct result of the smaller axial COM shifts in $D$.  In general, since the rotated endcap design $D$ has the smallest axial COM shifts it is expected to be the least susceptible to nonlinear resonance heating.


\begin{figure}
\includegraphics[width=85mm]{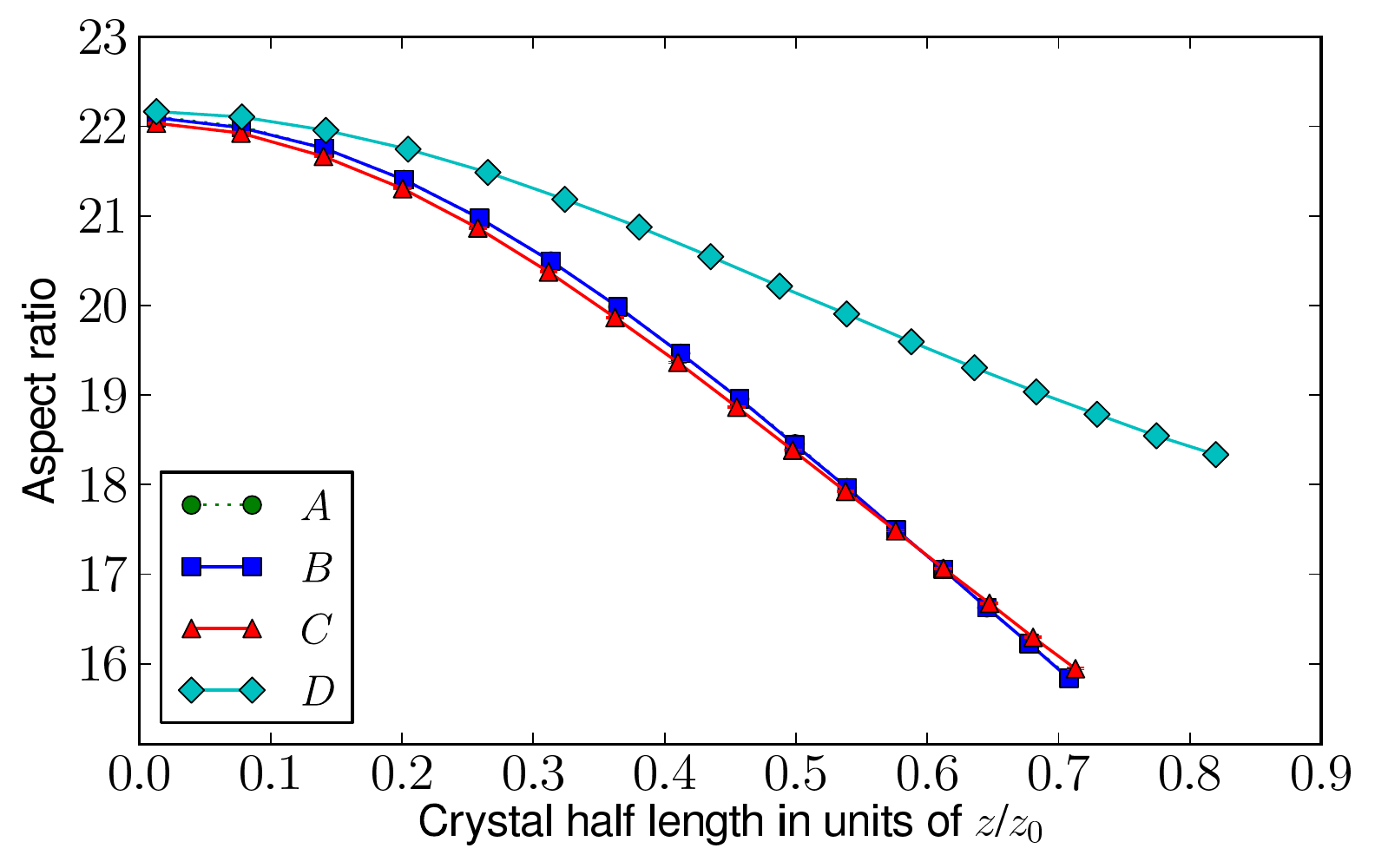}
\caption{{\label{fig-aspect_ratio}}(Color online.)  Aspect ratio as a function of crystal size, with aspect ratio estimated by comparison of the heights of the radial and axial potentials along $\hat{x}$ and $\hat{z}$ respectively.  $A$ is overlapped by $B$ and not visible.}
\end{figure}
The plots in Fig.~\ref{fig-wz} terminate at crystals of equal axial length.   For a perfectly harmonic trapping potential, the aspect ratio of a crystal will be independent of size and equal to the ratio of the secular frequencies, which in these traps were tuned to be equal in the harmonic trapping region.  Accordingly, eventual differences in aspect ratio are due to different degrees of anharmonicity sampled by the trapped ions.  In Fig.~\ref{fig-aspect_ratio}, we plot the expected change in aspect ratio during crystal growth in each design, calculated by a simple comparison of the heights of the radial and axial potentials along $\hat{x}$ and $\hat{z}$ respectively.  The tendency of design $D$ to maintain larger aspect ratio than other designs explains the ion numbers seen in Fig.~\ref{fig-wz}; at equal length, the crystals in designs $A$, $B$ and $C$ are radially wider than $D$ and so contain more ions.  Intuitively, the tendency of $D$ toward radially narrower crystals helps explain the smaller axial COM shifts in $D$, since the regions nearer the $\hat{z}$ axis are expected to be more harmonic than those radially further away.

The ripples in Fig.~\ref{fig-wz} caused by the finite size of the evaluation grid do not interfere with the conclusion that the COM frequency shift in $D$ is smaller than in the other designs, as the trend is clear at much sparser grids, where the ripples are even more pronounced.  The Laplace solutions are found to converge sufficiently to determine the COM shifts to within a few percent, and all other parameters--the associated Legendre and polynomial series fitting orders and the evaluation grid density--are adjusted until the resulting convergence in COM frequency shifts is not limiting.

\section{Conclusion}
\label{sec-conclusion}

We have compared the suitability of four trap designs for experiments on large Coulomb crystals.  The comparison is based on the desire to minimize micromotion heating and anharmonicity-induced nonlinear resonance heating.
 
We have compared total micromotion as a function of position within the trap volume. Micromotion heating, related to the total micromotion amplitude, is expected to be comparable in all designs.

We quantify axial micromotion by RMS averaging over a large trapping volume, finding that the resulting limit on spectroscopic accuracy in $B$ and $D$ to be much better than in either $A$ or $C$, with $C$ expected to be least suitable of all for precision spectroscopy on large crystals.

In order to compare the susceptibility of the different trap designs to nonlinear resonance heating, we have used a non-deforming crystal approximation to map the anharmonicity-induced axial and radial COM frequency shifts as a function of ion number.  Shifts of COM frequencies, as well as other normal mode frequencies for which COM shifts set the scale, are of concern since a growing crystal in an anharmonic trap will eventually cross a nonlinear heating resonance, leading to instantaneously elevated temperatures and potentially halting further crystal growth.  We find that a novel rotated endcap trap has the smallest axial COM shifts and observe that the relative stability of aspect ratio with increasing crystal size in this trap makes the result intuitive.  We suggest that traps aiming to grow large Coulomb crystals ought to minimize axial COM shifts, and we provide a method for quantifying these shifts.

The analysis presented here is simple to implement and should be predictive of trap behavior, so it is especially useful for taking a conservative approach to designing a trap for experiments on crystals of a desired size.  However, our analysis is not capable of predicting precisely when crystal growth will be limited by micromotion or by a given nonlinear heating resonance being crossed during crystal growth. To make detailed predictions of these limitations on ultimate crystal size and temperature, MD simulations could be used.  It would also be quite interesting to experimentally investigate the upper limits on crystal size set by anharmonicity-induced nonlinear heating resonances.

\begin{acknowledgments}
The authors gratefully thank Caroline Champenois, Eric Hudson, David Kielpinski, Joan Marler, Steven Schowalter, and Stephan Schiller for sharing their expertise in illuminating discussions.  This work is sponsored by NSF Grant No. PHY-0847748 and by NSF IGERT Grant No. 0801685.
\end{acknowledgments}

\bibliography{trappaperbib_v4}
\end{document}